\documentclass[12pt]{article}

\usepackage{scicite}
\usepackage{times}

\usepackage{authblk}
\usepackage[utf8]{inputenc}
\usepackage{graphics}
\usepackage{color}
\usepackage{caption}
\usepackage[demo]{graphicx}
\usepackage{subfig}
\usepackage[margin=1in]{geometry}
\usepackage{setspace}
\usepackage{subfig}
\usepackage[english]{babel}

\usepackage[backend=bibtex]{biblatex}
\usepackage{makecell}
\usepackage[svgnames]{xcolor}
\usepackage{hyperref}
\hypersetup{%
	colorlinks= true,
	linkcolor = MidnightBlue,
	urlcolor  = MidnightBlue,
	citecolor= MidnightBlue
}

\title{The Effectiveness of Digital Interventions \\ on COVID-19 Attitudes and Beliefs}
\author
{Susan Athey,$^{1}$ Kristen Grabarz,$^{2}$ Michael Luca,$^{3}$ Nils Wernerfelt,$^{4}$\\

\normalsize{$^{1}$Stanford Graduate School of Business, 655 Knight Way, Stanford, CA 94305 USA (\href{mailto:athey@stanford.edu}{athey@stanford.edu})}\\
\normalsize{$^{2}$Harvard Business School, Soldiers Field, Boston, MA 02163 USA (\href{mailto:grabarz@hbs.edu}{grabarz@hbs.edu})}\\
\normalsize{$^{3}$Harvard Business School, Soldiers Field, Boston, MA 02163 USA (\href{mailto:mluca@hbs.edu}{mluca@hbs.edu})}\\
\normalsize{$^{4}$Meta Platforms, One Hacker Way Menlo Park, CA 94025 USA (\href{mailto:nilsw@fb.com}{nilsw@fb.com})}
}

\date{}

\begin{document}

\maketitle
\begin{abstract}
During the course of the COVID-19 pandemic, a common strategy for public health organizations around the world has been to launch interventions via advertising campaigns on social media. Despite this ubiquity, little has been known about their average effectiveness. We conduct a large-scale program evaluation of campaigns from 174 public health organizations on Facebook and Instagram that collectively reached 2.1 billion individuals and cost around \$40 million. We report the results of 819 randomized experiments that measured the impact of these campaigns across standardized, survey-based outcomes. We find on average these campaigns are effective at influencing self-reported beliefs, shifting opinions close to 1\% at baseline with a cost per influenced person of about \$3.41. There is further evidence that campaigns are especially effective at influencing users' knowledge of how to get vaccines. Our results represent, to the best of our knowledge, the largest set of online public health interventions analyzed to date. 
\end{abstract}

\singlespacing \noindent {\bf Significance Statement:} This paper studies the question of whether social media advertising campaigns can be a cost-effective tool to influence attitudes and beliefs about COVID-19 vaccines.  We conduct a meta-analysis of more than 800 social media advertising experiments conducted by 174 public health-related organizations, where each experiment was designed to estimate the effect of advertisements on COVID-19-related attitudes and beliefs. The experiments individually have too few subjects to reliably detect small effects of advertising. By pooling the data we obtain more precise estimates of the overall average effect of the campaigns.  The estimated average cost per influenced person, about \$3.41, can be compared to the social benefit when evaluating the cost-effectiveness of public health campaigns, and successful tactics can be prioritized for testing in future experiments.

\clearpage
\pagenumbering{arabic}
\section*{Introduction}

Throughout the rapidly evolving COVID-19 pandemic, policymakers and public health agencies have needed to communicate with citizens about mitigation measures ranging from mask wearing and social distancing to vaccines. Advertising on social media has emerged as a popular channel to quickly reach large numbers of people, and it has been used by public health organizations in nearly every country both to convey information and influence behavior. An understanding of the expected impact of these campaigns is important as such organizations continue to engage in interventions as the pandemic unfolds. Assessing these campaigns is further valuable as digital public health interventions become increasingly used to address broader health-related outcomes.

To speak to these questions, this paper aims to evaluate the impact of social media advertisements on a variety of COVID-19-related outcomes. Analyzing advertising campaigns run on Facebook and Instagram by 174 public health organizations around the world, we investigate three main questions. First, what effect did these social media advertising campaigns have? Second, how cost effective were they? Third, which types of outcomes have the campaigns been most effective at influencing? 

The campaigns in our sample were run between December 2020 and November 2021, reached users in nearly every country, and in aggregate consist of \$39.4 million dollars of ad spend. Without disclosing their identities, they include a wide range of public health organizations that span major multinational nonprofits, public health ministries, and local NGOs. To our knowledge, the dataset we analyze is the largest set of online public health interventions studied to date. 

The data have two key features relevant for our analysis. First, we have data from a large number of the campaigns that conducted experiments where exposure to the ads was randomized at the user level, allowing us to assess the causal effect of each campaign. This is especially important in the context of online advertisements, where selection bias is a significant obstacle in non-experimental data \cite{gordon2019comparison,eckles2018field,blake2015ebay, moshary2021and}. Randomized experiments have become more common in online advertising, and companies (including Meta) have developed standardized experimentation tools to facilitate testing. The campaigns we analyze all used these tools to conduct experiments. 

Second, we are able to combine the experiments with user-level survey data for a subset of users. The surveys ask a variety of questions - namely, a user's willingness to get a COVID-19 vaccine, belief in the importance of vaccination, belief in vaccine effectiveness, belief in vaccine safety, whether the advertiser is a trustworthy source of COVID-19 information, how knowledgeable the user feels about how to get the vaccines, and whether they think vaccines are socially acceptable.\footnote{In this paper, we will refer to these questions shorthand as Willingness, Importance, Effectiveness, Safety, Trustworthy Source, Knowledgeable, and Social Norms, respectively.} While not all advertisers asked all questions, the survey questions were largely standardized across campaigns, facilitating comparisons. As is common practice with such experiments, we classify responses into a binary outcome according to whether respondents were engaging in the public health behavior of interest (for instance, intending to get the vaccine) or had the relevant public health information (for instance, knowing where to get the vaccine). Looking across all studies, we can then see whether interventions had an impact on the binary outcome of interest.\footnote{The wording of each question, possible responses, and how answers were classified are all listed in the SM.} This approach is similar in spirit to the approach taken by other papers that aggregate a set of experiments with distinct outcomes.\footnote{For instance, \cite{dellavigna2022nudge} aggregates a variety of distinct binary outcomes, corresponding to whether an action was taken or not, from a large set of behavioral experiments. Examples include whether or not someone filled out a government form or whether or not someone paid a fine. Two other studies that similarly use meta-analytic methods to combine different treatments and outcomes are \cite{benartzi2017nudge,hummel2019nudge}.}

Overall, our combined findings suggest that these campaigns were effective at influencing peoples' attitudes and beliefs about the vaccine. We find an average increase in the fraction of positive responses of 0.55 percentage points ($p$ = 2e-13) across all experiments, with a baseline 55.7\% positive rate. While this point estimate is small on a per person basis, the reach of the campaigns implies that even under conservative assumptions, around 11.6 million individuals were influenced by these campaigns alone, at a cost of about \$3.41 per incremental person. Such campaigns provide an easily scalable intervention that can in aggregate shape the public health outcomes of a large number of citizens.\footnote{This scalability includes not just reaching more people but also reaching individuals who may be hard or costly to reach via other means. This may be particularly important for some subpopulations where there is evidence of a disproportionate impact of the pandemic \cite{alsan2021great}.}

These results can be broadly compared to those from other initiatives aimed at influencing vaccination decisions.\footnote{A separate branch of literature has evaluated impacts of interventions on other COVID-related behaviors. For example, \cite{chen2020causal} use smartphone data from 10 million devices and find large effects of stay-at-home orders on both movement and transmission rates.}  \cite{barber2021conditional} and \cite{sehgal2021impact} estimate a \$68 and \$49 cost per incremental COVID vaccination in Ohio from the Vax-a-Million lottery (though see \cite{walkey2021lottery,thirumurthy2021association}); in a separate study, \cite{campos2021monetary} find a cost per incremental vaccination on the order of \$400; and \cite{krieger2000increasing} estimate costs of \$88-\$380 for incremental flu vaccinations in seniors in the US. \cite{larsenCOVID} finds a much lower cost, specifically \$1 per incremental COVID vaccination from a location-randomized YouTube advertising experiment in select counties in the US.   \cite{who2021} explores methods of effective communication for influencing health outcomes, including through randomized experiments run on Meta. Since our data does not include information about actual vaccination decisions, it is hard to directly contrast our estimates to those. However, such estimates highlight how challenging it can be to influence health behaviors, and the potential value of identifying low cost, scalable interventions. When considering such interventions to influence vaccine uptake, there has been much advocacy for behaviorally-informed promotions \cite{bavel2020using,volpp2021covid}. The campaigns we analyze broadly fit into this category.

Our findings also connect to the literature on health nudges, which similarly tends to focus on low cost, scalable interventions. Much of this literature has focused on text-based interventions, which have shown potential across a number of domains, ranging from flu appointments to court appearances \cite{milkman2021text,buttenheim2022flu,fishbane2020behavioral}. Specifically in the context of COVID-19, \cite{dai2021behavioural} sent participants in California text-based reminders to make vaccination salient and easy to remember. They find that reminders sent one day and eight days after notification of vaccine eligibility increased vaccination rates by 3.57 percentage points and 1.06 percentage points, respectively. \cite{banerjee2020messages} analyze the effect of a video message randomly distributed via SMS to millions of individuals in West Bengal, India; they find substantial effects on both the treated individuals as well as non-treated community members on a broad range of COVID-related outcomes. One advantage of text-based interventions is that they may be more salient and thus have a larger effect relative to our effects. On the other hand, advertising campaigns do not require gathering phone numbers and may thus be more easily scaled to a large population.\footnote{A related vein of literature has focused on identifying mechanisms for effective communication around COVID that could then be implemented at scale. For example, holding a wide range of factors constant, \cite{alsan2021experimental} vary characteristics of the messenger and signal content in a video infomercial and find evidence of substantial heterogeneity in effectiveness only by shifting those attributes. Similarly, \cite{jordan2021don} find evidence that prosocial framings are important for shifting COVID-related outcomes across a range of experiments. Insights from studies such as these two could help inform both digital and non-digital interventions.} 

Our results can be related to prior large-scale meta-analyses of online advertising. The literature has highlighted major challenges due to low statistical power \cite{lewis2015unfavorable}. Meta-analysis is a natural way to address this challenge, but it requires access to data from the experiments of many advertisers and also creates challenges comparing effectiveness across heterogeneous advertising objectives. We are aware of only three other meta-analyses of digital advertisements that have comparable scale to our study.\footnote{\cite{johnson2017online} provides an overview of the literature, including related meta-analyses in other forms of advertising such as online search advertising and television; see also \cite{bart2014products} who analyze 54 mobile advertising campaigns, and \cite{huang2020social} who study social ads across 74 products.} First, \cite{johnson2017online} used internal data from Google's display advertising platform to study the effect of digital advertisements on website visits for 432 digital advertising campaigns, finding effects of 8\% of baseline website visits. Second, \cite{goldfarb2011privacy} analyzed 2,892 experiments carried out by a brand research firm, each using a similar survey methodology and sample size to the experiments considered here, finding an effect of about 10\% of the baseline on survey responses concerning intention to purchase.  Third, \cite{gordon2022close} analyze more than 600 advertising experiments on Facebook, comparing the estimated effects for different measures along what is referred to as a ``funnel'' or a customer's journey to a final action of interest. They find effects of 28\%, 19\%, and 6\% for measured outcomes that capture consumer behavior at the top, middle, and lower parts of the funnel. Our paper is the first we are aware of to conduct a large-scale meta-analysis of online public health campaigns across multiple outcomes. Our estimated effects (about 1\% over baseline) are substantially smaller than the effects found in these studies, suggesting that it is more difficult to change attitudes and beliefs about vaccination than it is to increase more standard advertising outcomes. Similar to \cite{gordon2022close}, we find smaller effects for outcome measures that are closer to the ultimate outcome of interest, vaccination. 

Building on our main results, we next look across the different survey outcomes to see which ones are most impacted by the campaigns. We find significant effects on Knowledge, Safety, Social Norms, and Importance (all have $p<$ 0.001), while no significant effects on Willingness, Effectiveness, or Trustworthy Source. There were individual campaigns that were able to significantly move these last three metrics, but we could not detect an overall average effect. Finally, we find evidence that the campaigns may have been particularly effective at shifting users' knowledge around the vaccines. Knowledge has the largest treatment effect point estimate (1.23 percentage points, $p$ = 5e-7), and it is significantly higher than nearly all the other coefficients.\footnote{In a two sided t-test, the coefficient for Knowledge is significantly greater than that for Effectiveness ($p$ = 0.038), Importance ($p$ = 0.003), Safety ($p$ = 0.034), Trustworthy Source ($p$ = 0.002), Willingness ($p$ = 0.020), and our overall estimate ($p$ = 0.008), including when dropping Knowledge studies ($p$ = 0.006). It is not significantly different from our estimate for Social Norms, though it is close in a one sided test ($p$ = 0.109).} We interpret this as suggesting digital advertising may be a particularly cost-effective channel for information dissemination. In contrast, many of the other outcome metrics relied more on persuasion, which may be a harder needle to move with users.

Finally, we conclude with examples of campaigns that were particularly successful. Given the complexity of the characteristic space of the ads, any rigorous meta-analysis, even at our sample size, is under-powered. Hence, we simply highlight descriptive statistics about these campaigns as well as which of a set of previously documented best advertising practices they follow.

Overall, our results suggest that social media advertising campaigns can be an important component of public health initiatives. Over the course of the past two years, health oriented organizations have engaged in a wide range of tactics in an effort to shift attitudes and behaviors\footnote{See \texttt{https://www.nga.org/center/publications/covid-19-vaccine-incentives/} for a list of different incentives offered in the U.S. alone, ranging from Girl Scout cookies to laps on a NASCAR track.}; key challenges with many of these include scalability, measurement, and generalizability. Digital advertising can help overcome these challenges. However, the small per person impact highlights that these campaigns are best thought of as part of a broader set of strategies. To this end, our paper complements the growing literature on designing effective public health interventions on COVID. 

The rest of our paper is organized as follows. The Data section describes our sample, the outcome variables, and the approach we use to analyze the campaigns; the Results section presents findings in greater detail; and the Conclusion summarizes. Additional analyses are provided in the Supplementary Materials (SM).

\section*{Data}

\subsection*{Overview}

We analyze a set of 819 randomized experiments that were conducted between December, 2020 and November, 2021. The experiments in our sample are derived from 376 distinct advertising campaigns and 174 organizations.  There are often multiple experiments associated within a single ad campaign, where each experiment corresponds to a specific survey outcome. For example, an advertiser may take one campaign and run three separate experiments that measure the impact of the campaign on Willingness, Importance, and Effectiveness. The average campaign in our data ran slightly more than two experiments; in other words, advertisers measured the impact of their campaigns on an average of about two outcome metrics each.\footnote{The platform normally caps the number of questions per campaign at three.}

The studies were all conducted using Meta's infrastructure for conducting advertising effectiveness experiments across Facebook and Instagram; through the rest of this paper, we refer to this infrastructure as ``the platform.'' We focus specifically on experiments that measure the extent to which advertisements affect individuals' attitude or beliefs as measured by survey questions.\footnote{These experiments are known as `Brand Lift Studies' in advertiser-facing documentation. There are many companies that offer Brand Lift experiments to advertisers, each with slightly different implementations and methodologies. These studies are commonly used to measure effects on outcomes such as ad recall, brand sentiment, or intent to purchase, but have become popular during the pandemic to also look at health related outcomes that may not be observable in log data.} We note that not all advertisers run these experiments, so our results are underestimates of the total impact of digital advertising interventions on Facebook and Instagram.\footnote{Meta imposes minimum budgets to run one of these studies that vary across country; for example, in the U.S. it is currently \$30,000, which is more than many advertisers' budgets.}

We limit the set of experiments to those measuring outcomes in one of seven categories mentioned earlier (see SM for details).\footnote{While the platform proposed standardized questions to the advertisers, they did have autonomy to adjust the language for the questions if they wished, so that we see some heterogeneity of questions asked within the seven categories we study. In our data there are no instances where the same campaign ran multiple experiments that each asked the exact same survey question. However, 38 campaigns ran multiple experiments measuring the same outcome variable using distinct questions. For example, a campaign may have run two separate experiments that both measured its impact on Knowledge, with different survey questions, such as \emph{Do you know your order of priority to get the COVID-19 vaccine?} and \emph{Do you know where to go to get a COVID-19 vaccine for yourself?} Omitting these campaigns yields no material shift in our results, and in the SM we conduct additional analyses that factor in this within-campaign heterogeneity.} Though not an exhaustive set of COVID-19 ad experiments, these seven categories were selected because they are the most prevalent across COVID-19 vaccine related experiments, with at least 40 experiments in each category. We use all studies that asked these questions and, following the platform's policy\footnote{See \texttt{https://www.facebook.com/business/help/2396060560411130}.}, restrict to users aged 18 and older.

The campaigns we study total \$39.4 million in ad spend, with a reach of 2.1 billion unique users translated across 15 languages. The average campaign cost more than \$100,000 and reached nearly 13 million people; these were substantive efforts, but importantly also not beyond the budget of what many public health organizations could conceivably spend on similar campaigns in the future. Table 1 provides summary statistics.

\begin{table}[htbp]
\caption*{Table 1. Summary statistics by outcome metric.}
\resizebox{\columnwidth}{!}{\begin{tabular}{l c c c c c c c c}
\hline
Category & Effectiveness & Importance & Knowledge & Safety & \makecell{Social \\Norms} & \makecell{Trustworthy \\ Source} & Willingness & Overall\\[0.1ex] 
\hline\hline
\\
\# Experiments & 64 & 237 & 94 & 218 & 107 & 59 & 40 & 819\\
\\
\# Unique Campaigns & 48 & 234 & 73 & 218 & 107 & 59 & 40 & 376\\
\\
\# Unique Organizations & 32 & 109 & 57 & 100 & 50 & 30 & 17 & 174\\
\\

Earliest Campaign Start & 2021-02-17 & 2020-12-08 & 2021-02-19 & 2020-12-15 & 2021-02-03 & 2021-02-24 & 2020-12-08 & 2020-12-08\\
\\
Latest Campaign End & 2021-11-15 & 2021-11-12 & 2021-11-03 & 2021-11-15 & 2021-11-14 & 2021-10-21 & 2021-06-15 & 2021-11-15\\
\\
Avg. \# People Reached & 17,932,855 & 12,833,435 & 15,341,751 & 11,959,068 & 12,497,147 & 9,510,847 & 27,297,621 & 12,913,047\\
                & (3,216,101) & (1,353,134) & (2,957,774) & (1,746,021) & (2,396,430) & (1,971,257) & (6,773,069) & (1,225,276)\\
                \\
Avg. \# Survey Resp per Experiment & 1,753.53 & 2,130.65 & 1,734.60 & 1,931.67 & 1,510.58 & 1,762.17 & 1,429.00 & 1,860.94\\
                & (151.09) & (90.27) & (128.96) & (90.81) & (120.24) & (148.21) & (155.72) & (45.75)\\
                \\
Avg. Campaign Cost & \$136,807 & \$65,487 & \$122,796 & \$132,491 & \$63,556 & \$104,177 & \$392,678 & \$105,183 \\
          & (\$41,543) & (\$6,376) & (\$36,000) & (\$26,694) & (\$13,409) & (\$24,353) & (\$142,627) & (\$16,872)\\
\\
\# Experiments Rejecting No Effect (0.1) & 7 & 23 & 22 & 29 & 14 & 3 & 1 & 99\\
\\
\# Experiments Rejecting No Effect (0.05) & 5 & 16 & 17 & 20 & 8 & 3 & 1 & 70\\
\\
Implied False Discovery Rate (0.1) & 0.914 & 1.000 & 0.427 & 0.752 & 0.764 & 1.000 & 1.000 & 0.827\\
\\
Implied False Discovery Rate (.05) & 0.640 & 0.741 & 0.276 & 0.545 & 0.669 & 0.983 & 1.000 & 0.585\\
\\
\# FDR Survivor Experiments (10\% FDR) & 2 & 14 & 5 & 5 & 1 & 2 & 0 & 27\\
\\
\hline
\end{tabular}}
\footnotesize{Note: Standard errors in parentheses.\# Experiments Rejecting No Effect (0.1) references the number of studies that are significant at the 0.1 level in a two-tailed t-test against the null of no treatment effect. Implied False Discovery Rate (0.1) estimates the false discovery rate if we accepted all experiments that were significant at the 0.1 level, as $FDR(0.1) = (0.1*n_e)/(n_{rej})$, where 0.1 is the level of significance, $n_e$ is the number of experiments, and $n_{rej}$ is the number of experiments rejecting no effect at 0.1. \# FDR Survivor Experiments is the number of experiments determined via the Benjamini-Hochberg algorithm to survive a false discovery rate of 10\% \cite{benjamini}.}
\end{table}

Although the campaigns we study reached billions of users, as we can see in Table 1, we observe a much smaller number of survey responses. The platform provides the experimentation service to enable advertisers to estimate the incremental impact of their campaigns on survey-based outcomes; however, the number of responses per experiment is limited by the platform. The limits are presumably motivated by the fact that users may only be willing to engage in a small number of surveys, and the user experience may be negatively impacted by too many surveys. Thus, the platform caps the number of respondents per study to balance the tradeoff between statistical power and user experience.\footnote{See the SM for more details on how these experiments are implemented.} 

In total, our dataset incorporates 1.5 million total responses across all experiments. Per experiment, the number of responses ranges from 300 to 4507, split across test and control groups. This does not give us too much power for each individual experiment. If an experiment of average sample size (1861 in our sample) and average baseline positive response rate (55\%) were to be analyzed using a difference in means between treatment and control, the minimum detectable effect size with 80\% power and a 10\% significance level would be about 0.06, close to 10\% of the baseline. The average effects we find below are an order of magnitude lower than that.\footnote{In response to an early draft of this manuscript, the platform started granting more exceptions to the normal response caps on these experiments in order to deliver better powered results.} Indeed, we see only 99 out of the 819 experiments rejected the null at the 10\% level, just slightly greater than the 10\% of experiments that would be expected if there were no treatment effects. 

Only 27 experiments on their own are included in a set of experiments determined to have a 10\% false discovery rate using the Benjamini-Hochberg procedure \cite{benjamini}. We also report the implied false discovery rate for the set of experiments that are individually significant at the 5\% level, and we repeat this for the 10\% level. This exercise is motivated by the idea that an organization might choose to further scale a campaign after seeing a statistically significant impact. We see that if organizations used the 5\% threshold for scaling, for the Knowledge outcome the false discovery rate would be about 1/4, while for the Importance outcome, the rate would be about 3/4. Although such a false discovery rate might not be problematic, as it is unlikely the campaigns would be harmful and in aggregate these campaigns would be cost-effective, our findings as a whole suggest that we are not well powered to well identify individually effective campaigns. This motivates the approach we pursue in this paper of conducting a meta-analysis of hundreds of experiments together rather than seeking to identify individual campaigns with positive effects.

 In the SM, we provide graphs of the CDFs of p-values both overall and for each metric, where we can see that overall and particularly for the Knowledge outcome, the CDFs of p-values depart from the uniform distribution that would be expected if there were no effects.

\subsection*{Validity of Survey-Based Outcomes}

We now turn to consider the validity of the survey-based outcome measures, and in particular the extent to which they (or changes in them) do not capture changes in beliefs or knowledge. Privacy and legal constraints prevent advertisers from asking about or measuring some ultimate quantities of interest on platform (e.g., health or vaccination status), but the self-reported measures may still be meaningful.\footnote{\cite{breza2021effects} ran a location-randomized experiment where different regions were targeted with ad campaigns on Facebook. This experimental design allowed measurement and detection of significant effects on off platform outcomes (namely, travel and actual COVID cases). This result, though from a single experiment, demonstrates the potential for relevant offline effects from similar digital ads.}  Here, we discuss two categories of potential concerns about these measures.

A first category of concerns relates to whether the survey outcomes as entered in the platform reflect beliefs and behaviors in the physical world. There are several considerations. Following established practice for social media brand campaigns, survey outcomes are a primary outcome that public health organizations have been using to evaluate their campaigns. Campaigns start, stop, and change based on the results of these experiments, dictating how entire ad budgets of COVID interventions are spent. Hence, it is important to understand how these outcomes have responded to campaigns to date, and to add to the understanding of public health organizations whose individual experiments to date have been under-powered to detect small effects.

Relatedly, a common goal of public health organizations is to simply shift attitudes and beliefs. Akin to traditional advertisers where campaigns may target different levels of the conversion funnel, many of these advertisers are aiming to move awareness or basic beliefs, and may invest in complementary tactics to change behavior once beliefs have been influenced. To the extent that the implementation details and survey responses provide insight about awareness and beliefs, the campaign experiment outcomes are informative.

Finally, for social media campaigns in general, there is evidence that responses to platform surveys correlate reasonably well with behaviors of interest. \cite{moehring2021surfacing} find an $R^2$ of 0.83 in a regression of country-level vaccine uptake on self-reported vaccine status collected from a survey on Facebook. And \cite{astley2021global} find correlation between survey metrics on Facebook and off-platform COVID-19 cases (see \cite{bradley2021we} and \cite{reinhart2021big} for further discussions and caveats). \cite{alekseev2020effects} finds a high degree of correlation between characteristics of businesses that Facebook users self-report to own and offline statistics from the US Census. Although the contexts and analyses from these studies are different, together they suggest that there does appear to be informative signal in social media survey outcomes.

A second category of concerns about survey results relates to whether the differences in survey outcomes between treated and control groups can be interpreted as the causal effect of the advertisements \cite{gordon2019comparison}. One issue is that the there can be systematic differences between the treatment and the control group due to the implementation of the randomized advertising experiment. Details are provided in the SM, but in short, randomization of assignment to ads takes place just before an ad was intended to be shown to a user, so that whether a user sees an ad is random within the experiment. However, after seeing an ad, whether a user in the treatment group is subsequently shown a survey depends on an additional factor that is not present for the control group. In particular, if after randomization into the treatment group, the platform intends to show the user an ad but the user scrolls past it or does not scroll to it at all, the user will not be sent a survey. This is done to only capture survey responses from users in the treatment group who actually saw the ad, but to the extent to which this behavior is correlated with the survey outcome, it could lead to confounding and thus biased estimates of treatment effects. 

In addition, even if the set of users who were sent surveys was perfectly randomized within each experiment, there still is the potential for differential survey response between the treatment and the control group. This might occur if individuals influenced by the ads were more likely to respond to the survey. We address both of these issues in the SM, where we show that along several observable dimensions, the treatment and control group are similar. We further address these concerns by adjusting for several observable characteristics of individual respondents in our analysis, as described below.

A final concern is that the population answering the surveys differs from the overall target population along unobservables correlated with our outcome variable. For example, certain age groups may be more likely to answer the surveys. Post-stratifying our results by age and gender (as we do) is an industry standard approach to address this concern; in addition, in the SM we also compare observables from a post-stratified sample with those from the target population and find reasonable overlap.

\section*{Results}

We now turn to our main results, which we break into three categories. First, we describe for each survey outcome, the average effect across all experiments that focused on that outcome. Second, we combine the treatment effect estimates with data about the number of unique people who received advertisements as well as the total cost of the campaigns to estimate how many people have been influenced and what the cost per influenced person is. Finally, we discuss characteristics of successful campaigns. \\

\noindent {\bf Meta-Analysis of Experiments} 

We begin by analyzing each experiment separately using the following weighted linear model:

\begin{equation}
response_{i} = (\tilde{X}_i'\beta + \beta_0)W_i + \tilde{X}_i'\gamma+ \varepsilon_{i} 
\end{equation}

\noindent where $response_{i}$ is an indicator for whether individual $i$ gave a positive response, $\tilde{X}_{i}$ is a matrix of de-meaned controls that could potentially be related to outcomes (age bucket, gender, expected click through rate, and expected conversion rate), and $W_{i}$ is a dummy variable denoting whether $i$ was in the treatment or control group. The expected click through and conversion rates are platform-generated estimates of how likely a user is to click on and complete the survey; age buckets are 18-24, 25-34, 35-44, 45-54, 55-64, and 65+. We de-mean and interact our covariates with the treatment indicator so that $\hat{\beta}_0$ remains an unbiased and consistent estimate of the average treatment effect even in the presence of heterogeneous treatment effects by our covariates \cite{imbens2015causal}.

In our regression, each response is weighted with post-stratification weights by age bucket and gender within the treatment and control group, such that both arms of each experiment are representative of the population reached by the relevant campaign. That is, we obtain the proportion of users in each age bracket and gender group reached by the campaign associated with a given experiment and divide it by the proportion of responses to obtain the weight for each response.\footnote{To reduce variance, they are trimmed to an upper bound of 3 and a lower bound of 0.3 to decrease the influence of outlying observations. Rerunning without any trimming yields no material difference in the results.} We post-stratify by these two variables as age and gender are basic demographics that advertisers are both frequently interested in and where heterogeneous effects are often observed.\footnote{At the experiment level, we observe many significant positive and negative coefficients on our age and gender interaction terms; rerunning our meta-analyses on these coefficients yields insignificant average effects, however.} In the SM we explore robustness to different weighting schema and find no material difference in the results.

This procedure is based off the one that Meta uses to analyze results from these survey-based experiments for advertisers. Repeating this approach for each of the experiments in our dataset, we are left with 819 estimates of average treatment effects and standard errors. The next step in the analysis is to combine these point estimates into estimated effects by outcome metric and to generate an overall, combined estimate. We do this following standard meta-analytic methods \cite{borenstein2021introduction,hedges2014statistical}.

Specifically, to generate the average effect for each outcome and overall, we combine the respective experiments using inverse variance weighting (Table 2). This approach estimates a single, homogeneous effect per category while minimizing variance. We present this approach due to its simplicity and the fact that there is not evidence of heterogeneity across all outcomes; in the SM we report several alternative specifications that allow for greater heterogeneity across experiments and outcomes and find very similar results.

\begin{table}[htbp]
\caption*{Table 2. Meta-analysis of experiments by outcome.}
\resizebox{\columnwidth}{!}{\begin{tabular}{l c c c c c c c c}
\hline
Category 			& Effectiveness & Importance & Knowledge & Safety & \makecell{Social \\Norms} & \makecell{Trustworthy \\ Source} & Willingness & Overall\\[0.1ex] 
\hline\hline
\\
Treatment Coefficient 	& 0.0045 		& 0.0043*** 	& 0.0123*** 	& 0.0062*** 	& 0.0081*** 	& 0.0012 		& 0.0010 		& 0.0055***\\
                      			& (0.0029) 	& (0.0012) 	& (0.0025) 	& (0.0016) 	& (0.0024) 	& (0.0027) 	& (0.0042) 	& (0.0008)\\
\\
p-value 				& 0.114		& 0.0004 		& 5e-7 		& 8e-5 		& 0.0006 		& 0.639 		& 0.807 		& 2e-13\\
\\
Cost per Influenced Person & \$2.43 & \$2.41 & \$0.77 & \$3.20 & \$1.00 & \$11.79 & \$17.14 & \$3.41\\
\\
Baseline Positive Response Rate & 0.505 & 0.672 & 0.575 & 0.501 & 0.556 & 0.365 & 0.517 & 0.557 \\
\\
Treatment Effect as \% of Baseline & 0.89\% & 0.64\% & 2.14\% & 1.24\% & 1.46\% & 0.33\% & 0.19\% & 0.99\% \\
\\ 
{\bf Power Calculations (Approximate)}\\
\\
\hspace{.6cm} \emph{Minimum Detectable Effect} & 0.0071 & 0.0030 & 0.0061 & 0.0039 & 0.0059 & 0.0066 & 0.0104 & 0.0019\\
\\
\hspace{.6cm} \emph{Power to Detect Given Effect Size} & 0.474 & 0.973 & 1.000 & 0.989 & 0.962 & 0.120 & 0.081 & 1.00\\
\\
\hspace{.6cm} \emph{\# Experiments Needed for 80\% Power} & 159 & 115 & 23 & 87 & 57 & 1,660 & 4,133 & 94\\
\\
\hspace{.6cm} \emph{\# Survey Resp. per Exp for 80\% Power} & 4,349 & 1,036 & 426 & 770 & 801 & 49,567 & 147,635 & 214\\
\\
\hline
\end{tabular}}
\footnotesize{Note: Standard errors in parentheses. For each column, we consider the set of associated experiments and calculate the inverse-variance weighted average treatment effect (row 1). The Baseline \% Positive Response is an unweighted mean across all the relevant experiments; calculating it using fixed or random effects models changes the numbers only slightly. The Cost per Influenced Person for each subset is calculated using the spend and number of unique people reached across all campaigns in the relevant subset. Finally, we include power calculations based on the standard error of the treatment effects (abstracting away from heterogeneity across experiments). Power to detect a given effect size is calculated at the $\alpha=.1$ level; for the last two rows, we want to convey how power could be improved by either increasing the number of experiments or the number of surveys per experiment. For those calculations we ask how much of either we would need holding the other fixed to have 80\% power to detect the given estimated treatment effect. See supplemental code for exact details.}\\
\end{table}

Several comments are salient to the interpretation of these results. First, this dataset is very broad. Past efforts to understand what has and has not worked with shifting behaviors around COVID have often by necessity studied a small number of treatments at modest scale or been one off ex post analyses. External validity with such studies is frequently a concern and potentially helps explain why studies to date have found conflicting results (e.g., \cite{chang2021financial,campos2021monetary,thirumurthy2021association,walkey2021lottery,sehgal2021impact} on effects of financial incentives on vaccination rates). In contrast, here, pooling hundreds of studies, we find a positive and statistically significant average main effect. The evidence may not be conclusive yet on what kinds of behavioral nudges work, but this is evidence that these digital advertising campaigns can help move the needle on COVID-related attitudes. 

Second, consider results about the specific outcomes. We find that Importance, Knowledge, Safety, and Social Norms showed highly statistically significant effects. In contrast, we do not detect an effect for Effectiveness, Trustworthy Source, or Willingness. (Though Effectiveness is close to marginally significant in our main specification above.)

We note that these last three metrics, particularly Effectiveness and Willingness, are arguably lower in the vaccine conversion funnel (that is, they are better proxies for a final desired action) than the first four metrics. A stylized fact from advertising is that such ``lower-funnel'' behaviors often see smaller effect sizes than more upper level outcomes, and are generally challenging to study as they may be influenced by a variety of unmeasured factors. While our finding thus accords with this intuition, a limitation of our study is that even with a very large sample size, we are not powered to generate more precise estimates of these averages (see the final rows of Table 2).

Third, we note that there is evidence the estimated lift for Knowledge (1.23pp) is significantly higher than our other metrics. Our estimate for Knowledge is significantly greater than that for Effectiveness ($p$ = 0.038), Importance ($p$ = 0.003), Safety ($p$ = 0.034), Trustworthy Source ($p$ = 0.002), Willingness ($p$ = 0.020), and our overall estimate ($p$ = 0.008), including when dropping the Knowledge experiments ($p$ = 0.006). It is nearly significantly greater than Social Norms in a one sided t-test ($p$ = 0.109). We note that Knowledge is a distinct outcome here in that the other outcomes relate more to persuasion, whereas Knowledge focuses simply on conveying information. These results suggest that social media campaigns may be particularly attractive for public health organizations interested in the latter.\footnote{In thinking about the effects for Knowledge as well as the other metrics, a relevant data point is the baseline positive response rate for Trustworthy Source ($36.5$\%). While this question was only asked in a subset of campaigns it is still revealing that the effects we are seeing are despite a relatively low user trust level with some advertisers. This suggests that health organizations with strong brand values may be well positioned to see particularly large effects, a topic we leave to future research.}

Finally, we note that in the SM we explore different specifications and find broadly similar results. In addition, the standard output provided to advertisers on Meta comes from a Hierarchical Bayesian Model; we chose a frequentist approach due to its simplicity, but in the SM we show robustness to using a similar Bayesian approach. \\

\noindent {\bf Number of influenced people, cost per influenced person.} Conditional on our results, how many people were influenced by these campaigns, and how cost effective were they? As aforementioned, the survey data only comes from a subset of the overall users who saw the ads; to calculate the number of influenced people, we follow the common industry practice and scale the point estimate of the treatment effect from each experiment by the size of the overall population that saw the campaign. In our case, since we have data across many advertisers and some users were shown ads from multiple campaigns, to generate a (conservative) estimate of the number of influenced people per campaign, we can treat these collective campaigns as effectively one large one. Specifically, we combine the total spend, total unique reach, and our estimate of the average treatment effect to generate a back of the envelope estimate of the number of people who were influenced by this combined effort.

Doing this calculation, we estimate that about 11.6 million people were influenced by these campaigns. To be clear, by `influenced' we mean shifted self-reported beliefs to a positive outcome; this does not capture people who, for example, moved along the intensive margin of these categories.\footnote{In the SM we explore a less conservative way of estimating the number of influenced people (and thus the cost per influenced person).}

Conditional on estimates of how many people were influenced, how much does it cost to influence someone? For this, we divide the number of influenced people by the ad spend, again as is typical in the industry. From Table 2 we can see the average cost per influenced person was \$3.41. While we are hesitant to extrapolate substantially outside our sample, these magnitudes suggest running even a few million dollars' worth of additional campaigns could achieve relatively large shifts in the baseline fraction of positive responses.
\\

\noindent 
\noindent {\bf Examples and characteristics of successful campaigns.} Finally, what are some examples of successful campaigns? Given the many dimensions in which these advertising campaigns differ from one another, even with our sample size, any rigorous meta-analysis would be under-powered. Nonetheless, to lay the groundwork for future research, we provide examples of ads that saw significant lifts in their outcome metrics (Figure 1).

\begin{figure}[htp]
\centering
\includegraphics[width=\textwidth]{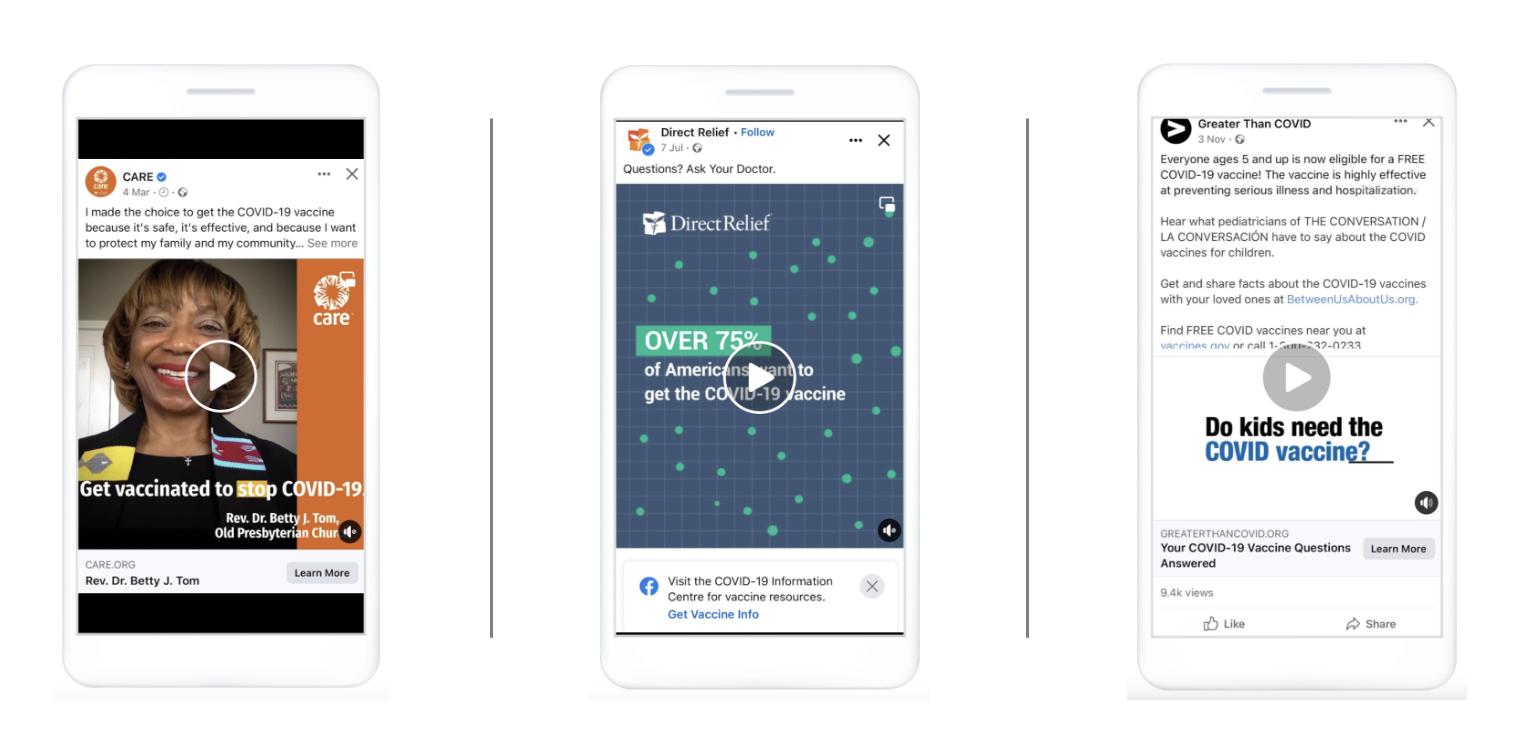}
\caption*{Figure 1. Examples of successful campaigns from CARE US, Direct Relief, and the Kaiser Family Foundation, respectively.}
\end{figure}

In discussing these examples, we highlight relevant best practices that Meta has previously publicly shared based on analysis of other historical data (Table 3).\footnote{Available at \texttt{https://www.facebook.com/business/help/370852930116232?id=271710926837064} and \texttt{https://www.facebook.com/gpa/blog/creative-considerations}. Paraphrased here.} \\

\begin{table}[!htbp] \centering 
  \caption*{Table 3. Best practices, creative and technical.} 
  \label{} 
\begin{tabular}{@{\extracolsep{5pt}} ll} 
\\[-3ex]\hline 
\hline \\
$\bullet$ Have a clear problem statement & $\bullet$ Use vertical videos \\ 
$\bullet$ Show a specific solution  & $\bullet$ Do not put excessive text over images \\ 
$\bullet$ Make it easy to act & $\bullet$ Keep text short\\ 
$\bullet$ Highlight the cause, not the organization & $\bullet$ Add multiple images using the carousel format \\ 
$\bullet$ Use a credible speaker & $\bullet$ Add movement \\ 
& $\bullet$ Use calls to action \\
\hline \\[-1.8ex] 
\end{tabular} 
\end{table}

\noindent Given the above, we now review each example to highlight some of these best practices:

\begin{itemize}
	\item {\bf CARE US}: This ad uses a faith leader within the target audience to bring a credible voice (CARE tested different messengers across campaigns to see which ones would resonate). The ad is actually a short video loop where the `stop COVID-19' gets highlighted, adding movement to the ad to help catch users' attentions while also drawing attention to the cause.
	\item {\bf Direct Relief}: Note the vertical aspect ratio of the animation and the concise and clear text. Further, the video features a series of dynamic messages with similar punchiness that ends with a clear call to action.
	\item {\bf Kaiser Family Foundation}: This ad features a clear problem statement, with multiple easy to follow calls to action. It further features a prominent physician within the local community as a credible voice.
\end{itemize}

\noindent These examples suggest characteristics of ads that might be analyzed in more detail in future research.

\section*{Conclusion}

Prior to COVID-19, many public health organizations had never run advertising campaigns on social media. However, over the course of the pandemic, there has been a dramatic increase in this type of campaign that not only shows no sign of stopping for the remainder of the pandemic, but has also opened the door for future digital interventions on other topics (e.g., childhood vaccination, handwashing). These factors highlight the importance of better understanding and quantifying the effects of these interventions.

Our results show that public health interventions via digital advertising are an effective medium for changing important self-reported beliefs and attitudes around COVID-19. Further, the cost-effectiveness and scale of these campaigns can make them appealing to a broad range of organizations around the world. However, we highlight the long-recognized problem that statistical power is a large problem for evaluating effectiveness, particularly for the measures that relate most closely to individual action \cite{lewis2015unfavorable}.

Finally, we end by noting how public health and advertising have traditionally been viewed as two distinct domains. A barrier to entry for many public health organizations in digital advertising has simply been a lack of familiarity, which in turn leads to skepticism that these campaigns could actually effect change. With more information about best practices, effect sizes, and cost-effectiveness, it may be possible to improve the quality of these campaigns and encourage their use in cases where they have the potential to be effective.

\pagebreak

\printbibliography

\subsection*{Acknowledgments}
We thank CARE US, Direct Relief, and the Kaiser Family Foundation for granting us permission to use their ads. We also thank Kang-Xing Jin and the health group at Meta Platforms for driving the creation of these public health interventions in the first place. S.A. acknowledges funding from the Golub Capital Social Impact Lab at Stanford Graduate School of Business.

\subsection*{Funding}
Funding to hire K.G. as a part-time contractor to work on this project came from Meta Platforms.

\subsection*{Author Contributions}
Author listing is alphabetical, and all participated significantly in the design, analysis, and writing.

\subsection*{Competing Interests}
N.W. is an employee of Meta Platforms and owns stock. K.G. is a part-time contractor through PRO Unlimited, a contracting agency used by Meta Platforms. S.A. received funding from Meta Platforms for research projects related to public health, and she previously provided consulting services to Meta. M.L. declares no competing interests.

\pagebreak

\section*{\centering{Supplementary Materials for}}
\subsection*{\centering{Effects of Digital Interventions \\ on COVID-19 Attitudes and Beliefs}}
\setcounter{page}{1}

\subsection*{\centering{Susan Athey,$^{1}$ Kristen Grabarz,$^{2}$ Michael Luca,$^{3}$ Nils Wernerfelt$^{4}$}}

\begin{centering}
\normalsize{$^{1}$Stanford Graduate School of Business, 655 Knight Way, Stanford, CA 94305 USA (\href{mailto:athey@stanford.edu}{athey@stanford.edu})}\\
\normalsize{$^{2}$Harvard Business School, Soldiers Field, Boston, MA 02163 USA (\href{mailto:grabarz@hbs.edu}{grabarz@hbs.edu})}\\
\normalsize{$^{3}$Harvard Business School, Soldiers Field, Boston, MA 02163 USA (\href{mailto:mluca@hbs.edu}{mluca@hbs.edu})}\\
\normalsize{$^{4}$Meta Platforms, One Hacker Way Menlo Park, CA 94025 USA (\href{mailto:nilsw@fb.com}{nilsw@fb.com})}\\
\end{centering}

\section*{Supplementary Text}

This section has six parts: (i) we provide an overview of the advertising experimentation infrastructure at Meta and how our studies were implemented; (ii) we run various robustness checks on our main analyses; (iii) we provide graphs of the CDFs of the $p$-values overall and by metric; (iv) we explore a less conservative way of calculating the cost-effectiveness of our campaigns; (v) we document characteristics of survey respondents across groups to provide insight into any possible bias; and (vi) we list the questions asked, possible responses, and bucketing into positive responses and not.

\subsection*{Overview of Advertising Campaign Experiments at Meta}

``Lift" studies are the core tool advertisers use to assess the impact of their campaigns on Meta. Within such studies, ``Brand Lift Studies" are those which focus on survey-based outcomes and are the ones we focus on in this paper. Meta has an external-facing self-service tool for Brand Lift Studies, though large enough advertisers have access to a liaison within Meta who can set up more sophisticated studies (e.g., ones that have multiple treatment arms, custom questions, and varying holdout sizes). Despite this heterogeneity in how advertisers interact with the platform, the technical implementation of the lift studies is standardized. This greatly facilitates analysis across studies.

Each time such a campaign experiment is created, all users are hashed into treatment and control for that campaign. When an auction is run for a specific user, if an ad from the study in question wins the auction and the user is in the treatment group, the ad is sent to the user, but if the user is in the control group, the ad is held back. Both treated and control users are eligible to be sent a poll 4-48 hours after the corresponding (potential) exposure event. Given a total desired number of responses, surveys are sent to users spaced out over the course of the campaign, where each user may only be polled once per study. Natural variation in response rates means responses are not always uniform across time within the course of the campaign, but in practice the variation tends to be minimal. 

A complication here is even if Meta's servers send a user an ad, that does not necessarily guarantee she will see it. The physical layout of the delivery surfaces themselves may mean, for example, that a user may be sent an ad but they may not scroll down far enough to see it, or may scroll too fast. In the experimental setup, only users in the treatment group who saw an ad are eligible to receive surveys. This avoids sending surveys to users assigned to test but who did not see the ad, which would capture an ITT advertisers arguably may care less about. In contrast, any user in the control group where the treatment ad won the auction and was held out from being sent is eligible to receive a survey, whether or not they saw the slot where the ad would have been. This ensures that no user who is sent a survey in the control group was exposed the test ad (at least on their account), but it does create a slight difference in the survey-eligible populations across treatment and control. For example, users who scroll quickly may be more likely to be sent surveys if they are in the control. In practice, the differences between the two populations has proven slight -- our results, as do those reported out to advertisers, contain controls for activity bias which is by far the largest source of differential exposure across groups, and even unconditional on that, internal analyses have shown that the two populations do not significantly differ along a wide range of observables. For completeness, however, we perform an additional check later in the SM.

Conditional on this data generating process, we have the data on respondents, their demographic characters, and those of the target population that we use in our main analyses. We note this is the same raw data that Meta analyzes and reports out to advertisers, though as aforementioned, we do our own, custom analysis.

\subsection*{Robustness of Results to Different Specifications and Weighting Schemes}

\noindent Below we report on different specifications for our main results. We run three broad categories of alternative analyses: (i) we vary the controls, weighting, and linear specification in our experiment-level analysis; (ii) we allow for heterogeneity in our treatment effect estimates across experiments (`random effects' meta-analysis methods); and (iii) we explore how our results differ under a Bayesian analysis. We present the respective results in turn. Unless otherwise noted, conditional on experiment-level results, estimates are combined using inverse variance weighting to generate the main point estimates.\\

\noindent {\bf Varying the controls, weights, and linearity}

\noindent Recall our main specification:

\begin{equation}
response_{i} = (\tilde{X}_i'\beta + \beta_0)W_i + \tilde{X}_i'\gamma+ \varepsilon_{i} 
\end{equation}

\noindent with variables described as in the main text. In particular, within $\tilde{X}_i$ we have platform-generated controls for expected click through and expected completion rates of the survey. One may be concerned that these are rather black box and may be correlated with our treatment variable in a way that may affect inference. Hence, we rerun our main results dropping these platform-generated controls.

Second, in our main specification, the individual responses are weighted based on their age and gender category and the frequency of those buckets in the campaign's target population, such that the treatment and control groups are both representative of the campaign demographics in each experiment. To explore the sensitivity of our results to this weighting, below we rerun our baseline model with different weights (only gender and only age). 

Finally, while we estimate the above using OLS, logistic regressions are commonly used in these settings (e.g., due to their bounded range), so we report results out using that as well. Specifically, for each experiment we estimate the average marginal effect of our treatment and then combine those across experiments. Across all these specifications, which are reported in Table S1, our results are broadly similar -- or stronger.\\

\begin{table}[htbp]
\caption*{Table S1. Varying the controls, weights, and linearity.}
\resizebox{\columnwidth}{!}{\begin{tabular}{l c c c c c c c c}
\hline
         & Effectiveness & Importance & Knowledge & Safety & \makecell{Social \\Norms} & \makecell{Trustworthy \\ Source} & Willingness & Overall \\[0.1ex] 
\hline\hline
\\
\multicolumn{9}{l}{\textbf{Main results}}\\
\\
Treatment Coefficient 	& 0.0045 		& 0.0043*** 	& 0.0123*** 	& 0.0062*** 	& 0.0081*** 	& 0.0012 		& 0.0010 		& 0.0055***\\
                      			& (0.0029) 	& (0.0012) 	& (0.0025) 	& (0.0016) 	& (0.0024) 	& (0.0027) 	& (0.0042) 	& (0.0008)\\
\\
p-value 				& 0.114		& 0.0004 		& 5e-7 		& 8e-5 		& 0.0006 		& 0.639 		& 0.807 		& 2e-13\\
\\ 
\hline
\\
\multicolumn{9}{l}{\textbf{Dropping platform-generated controls}}\\
\\
Treatment Coefficient & 0.0006 & 0.0032* 	& 0.0124*** 	& 0.0044* 	& 0.0076** 	& 0.0031 		& -0.0005 		& 0.0044***\\
        			& (0.0034) & (0.0015) 	& (0.0033) 	& (0.0017) 	& (0.0028) 	& (0.0030) 	&(0.0054)		& (0.0007) \\
\\
p-value		 & 0.858 	& 0.030 		& 2e-4 		& 0.011 		& 0.007 		& 0.303 		& 0.921 		& 	2e-9\\
\\
\hline
\\
\multicolumn{9}{l}{\textbf{Post stratification weighting by age bracket only}}\\
\\
Treatment Coefficient 	& 0.0034		& 0.0041** 	& 0.0112* 	& 0.0065*** 	& 0.0083** 	& 0.0028 		& 0.0012		& 0.0057***\\
        				& (0.0038) 	& (0.0016) 	& (0.0044) 	& (0.0018) 	& (0.0028) 	& (0.0031) 	& (0.0042)		& (0.0008) \\
\\
p-value 			& 0.364 		& 0.008 		& 0.011		& 0.0003 		& 0.003		& 0.367 		& 0.782		& 7e-14\\
\\
\hline
\\
\multicolumn{9}{l}{\textbf{Post stratification weighting by gender only}}\\
\\
Treatment Coefficient 	& 0.0050 		& 0.0045** 	& 0.0119** 	& 0.0065*** 	& 0.0079** 	& 0.0039 		& 0.0017		& 0.0060***\\
        				& (0.0035) 	& (0.0015) 	& (0.0042) 	& (0.0017) 	& (0.0027) 	& (0.0032) 	& (0.0040)		& (0.0007) \\
\\
p-value 			& 0.147		& 0.002 		& 0.005 		& 0.0002	 	& 0.004 		& 0.224 		& 0.683		& 2e-16\\
\\
\hline
\\
\multicolumn{9}{l}{\textbf{Logistic regression}}\\
\\
Average Marginal Effect 	& 0.0025		& 0.0056*** 		& 0.0141*** 		& 0.0052*** 	& 0.0089*** 	& 0.0061* 		& 0.0048		& 0.0064***\\
        				& (0.0030) 	& (0.0013) 		& (0.0026) 		& (0.0017) 	& (0.0026) 	& (0.0032) 	& (0.0057)		& (0.0008) \\
\\
p-value 			& 0.395 		& 3e-5 			& 4e-8			& 0.002 		& 0.001 		& 0.060		& 0.406		& 6e-15\\
\\
\hline \\
\end{tabular}}
\footnotesize{Note: Standard errors in parentheses.}\\
\end{table}
\pagebreak

\noindent {\bf Incorporating heterogeneous effects across experiments}

Conditional on our experiment-level results, the analysis in the main text combines them together using inverse variance weighting. This assumes a homogeneous treatment effect across all pooled studies (a `fixed effects' meta-analysis). In practice, however, we may think treatment effects may vary substantially across campaigns or advertisers, motivating an analysis that estimates a distribution of treatment effects (a `random effects' meta-analysis). We report the results in the main text given their simplicity and the fact that for some metrics the heterogeneity does not seem very substantial \cite{higgins2019cochrane}. Further, incorporating distributions of effect sizes, as we do below, the results do not change substantively. 

In particular, we run two separate analyses below, which are both reported in Table S2. First, we do a standard random effects meta-analysis within each outcome. Specifically, for experiment $i$ that asks question $q$ with observed effect size $\hat{\beta}_{qi}$ and observed standard error $\sigma_{qi}$, we allow the true effect size, $\beta_{iq}$ to vary around a true, grand average effect for that question, $\bar{\beta}_{q}$, with variance $\tau^2_q$. As is common in such models, we parametrize the true treatment effect for the experiment, $\beta_{iq}$, to be drawn from a normal distribution $N(\bar{\beta}_q, \tau^2_q)$ and then the observed treatment effect $\hat{\beta}_{qi}$ to be drawn from a normal $N(\beta_{iq}, \sigma^2_{iq})$. We do not want to lean on this too heavily, but we note past work has found such Gaussian assumptions in random effects meta-analyses can perform reasonably well even if the true data are non-Gaussian \cite{mcculloch2011misspecifying}. We fit this via maximum likelihood for each outcome separately, reporting the results for the grand mean and its corresponding prediction interval below. For the Overall column, we repeat this procedure but across all experiments.

Second, given the same campaign is often involved with multiple experiments, we can allow correlation within effects for each campaign. This is mainly an issue for estimating the Overall effect, though there are campaigns that did ask slight variants of the same question (as discussed in the main text). This constitutes a three-level random effects meta-analysis where we allow a distribution over campaign effects, a distribution within each campaign across questions, and then a final distribution over observed experiment effects conditional on the true treatment effect. We parametrize this similarly as three nested normals: for a campaign $i$ and question $q$, the distribution of true campaign effects, $\beta_i$, is $N(\bar{\beta},\tau^2_1)$; the distribution over question true effects, $\beta_{iq}$, within each campaign is given by $N(\beta_i, \tau^2_2)$; and the distribution of realized experiment treatment effects, $\hat{\beta}_{iq}$, is $N(\beta_{iq},\sigma^2_{iq})$. We again fit this via maximum likelihood for each outcome separately and then for Overall. 

Finally, we do not report it as it is only for the Overall outcome, but we also ran a three-level random effects model with a distribution over questions, a distribution over experiments within each question, and then a distribution over realized effects. This yields an estimate of 0.56pp, which is again significant from zero ($p<$ 0.0001).

Below we also include estimates of Cochran's $Q$ \cite{cochran1954some}, a common statistic in meta-analyses to test for heterogeneity across experiments. Cochran's $Q$ is a weighted sum of squares across experiments -- specifically, it is defined as $Q = \sum_{k} (\hat{\theta}_k - \hat{\theta})^2/\hat{\sigma}^2_k$ where the summation is taken over all experiments in question, $\hat{\theta}_k$ is the estimated treatment effect in study $k$, $\hat{\theta}$ is the estimated treatment effect of the studies derived via a fixed effects model, and $\hat{\sigma}^2_k$ is the estimated variance of the treatment effect estimate in study $k$. Note that intuitively higher $Q$ values mean there is greater dispersion in treatment effect estimates around the pooled estimate, evidence of heterogeneity. Cochran's $Q$ follows a Chi-squared distribution, with low p-values meaning a rejection of the null of homogeneity. While suggestive, this test in isolation should not be used to determine between fixed and random effects meta-analyses (see \cite{higgins2019cochrane} for more details).\\

\begin{table}[htbp]
\caption*{Table S2. Random effects models.}
\resizebox{\columnwidth}{!}{\begin{tabular}{l c c c c c c c c}
\hline
         & Effectiveness & Importance & Knowledge & Safety & \makecell{Social \\Norms} & \makecell{Trustworthy \\ Source} & Willingness & Overall \\[0.1ex] 
\hline\hline
\\
\multicolumn{9}{l}{\textbf{Main results}}\\
\\
Treatment Coefficient 	& 0.0045 		& 0.0043*** 	& 0.0123*** 	& 0.0062*** 	& 0.0081*** 	& 0.0012 		& 0.0010 		& 0.0055***\\
                      			& (0.0029) 	& (0.0012) 	& (0.0025) 	& (0.0016) 	& (0.0024) 	& (0.0027) 	& (0.0042) 	& (0.0008)\\
\\
p-value 				& 0.114		& 0.0004 		& 5e-7 		& 8e-5 		& 0.0006 		& 0.639 		& 0.807 		& 2e-13\\
\\ 
\hline
\\
\multicolumn{9}{l}{\textbf{Random effects within each outcome}}\\
\\
Treatment Coefficient 			& 0.0033 		& 0.0040*** & 0.0106** & 0.0065*** 				& 0.0077*** 		& 0.0030				& 0.0010 		& 0.0056***\\
        						& (0.0038) 	& (0.0015) & (0.0044) & (0.0018) 				& (0.0028) 		& (0.0032) 			& (0.0042)		& (0.0010) \\
\\
p-value 					& 0.382 		& 0.009 	& 0.015   & 0.0003 					& 0.006		& 0.337 					& 0.806		& 9e-9\\
\\ 
Cochran's $Q$ ($p$-value) 	& 96 (0.004)   	& 424 ($<$1e-4) &  232 ($<$1e-4)  & 412 ($<$1e-4) 	& 167 ($<$1e-4) 		& 113 ($<$1e-4) 		& 39 (0.468) 	& 1497 ($<$1e-4)\\
\\
\hline
\\
\multicolumn{9}{l}{\textbf{Three-level random effects}}\\
\\
Treatment Coefficient & 0.0027 	& 0.0043*** 	& 0.0118** 			& 0.0062*** 	& 0.0077*** 		& 0.0030 			& 0.0010		& 0.0052***\\
       			 & (0.0040) 	& (0.0012) 	& (0.0049) 			& (0.0016) 	& (0.0028) 		& (0.0032) 		& (0.0042)		& (0.0011) \\
\\
p-value 		& 0.493 		& 0.0004		& 0.015				& $<$1e-4		& 0.006        		& 0.337			&  0.807		& $<$1e-4\\
\\
Cochran's $Q$ & 96 (0.004) 	& 424 ($<$1e-4) & 232 ($<$1e-4) 		& 412 ($<$1e-4)  & 167 ($<$1e-4)	 & 113 ($<$1e-4) 		& 39 (0.468) 		 & 1516 ($<$1e-4)\\
\\
\hline \\
\end{tabular}}
\footnotesize{Note: Standard errors in parentheses.}\\
\end{table}
\pagebreak

\pagebreak

\noindent {\bf Bayesian Analysis}

Finally, we use a Bayesian approach to rerun one of our earlier specifications as a sanity check. Bayesian meta-analyses have become popular recently for a number of reasons (e.g., \cite{meager2019understanding,higgins2019cochrane}), but a general advantage in random effects settings is an ability to directly model the uncertainty in the between study variance. Hence, below we run a Bayesian version of the first random effects model we outlined above to see if there is any evidence the Bayesian approach produces substantively different estimates.

Specifically, we assume a hierarchical likelihood as follows. For all experiments $i$, with estimated treatment effect $\hat{\beta}_i$ and estimated standard error $\hat{\sigma}_i$, we assume $\hat{\beta}_i \sim N(\beta_i, \hat{\sigma}^2_i)$, where $\beta_i \sim N(\beta, \tau^2)$. Following the defaults of \cite{meagerbaggr}, we specify weak priors of $\beta\sim N(0,100)$ and $\tau \sim U[0,10\bar{\sigma}]$ where $\bar{\sigma}$ is simplistically defined as the standard deviation of the estimated $\hat{\beta}_i$'s. Given this set up, we then run the Bayesian model for all of the experiments that correspond to each outcome as well as all of them collectively for the Overall estimate. In Table S3 we report estimates of the hypermeans and corresponding uncertainty intervals; we can see that compared to the above, there is reassuringly little difference in the estimates.\\

\begin{table}[htbp]
\caption*{Table S3. Bayesian analysis.}
\resizebox{\columnwidth}{!}{\begin{tabular}{l c c c c c c c c}
\hline
         				& Effectiveness & Importance & Knowledge & Safety & \makecell{Social \\Norms} & \makecell{Trustworthy \\ Source} & Willingness & Overall \\[0.1ex] 
\hline\hline
\\
\multicolumn{9}{l}{\textbf{Main results}}\\
\\
Treatment Coefficient 	& 0.0045 		& 0.0043*** 	& 0.0123*** 	& 0.0062*** 	& 0.0081*** 	& 0.0012 		& 0.0010 		& 0.0055***\\
                      			& (0.0029) 	& (0.0012) 	& (0.0025) 	& (0.0016) 	& (0.0024) 	& (0.0027) 	& (0.0042) 	& (0.0008)\\
\\
p-value 				& 0.114		& 0.0004 		& 5e-7 		& 8e-5 		& 0.0006 		& 0.639 		& 0.807 		& 2e-13\\
\\ 
\hline
\\
\multicolumn{9}{l}{\textbf{Bayesian Analysis.}}\\
\\
Hypermean Estimate	 & 0.0032			& 0.0040** 			& 0.0106** 		& 0.0065**		& 0.0077** 		& 0.0030			& 0.0012 				& 0.0056**\\
        				& [-0.0048, 0.0108] 	& [0.0010, 0.0071] 	& [0.0015, 0.0190]	& [0.0030, 0.0101]	& [0.0023, 0.0132] 	& [-0.0032 0.0100] 	&  [-0.0078, 0.0103]		& [0.0037, 0.0075]	 \\
\\
\hline\\
\end{tabular}}
\footnotesize{Note: 95\% uncertainty intervals in parentheses. Two asterisks denotes 0 is not contained in the 95\% uncertainty interval.}\\
\end{table}

\pagebreak
\subsection*{CDFs of p-values}

To further explore evidence on the potential rate of false positives in our study, in Figures S1 and S2 we plot CDFs of the p-values across studies, both overall and for each metric. For both the overall distribution of p-values and the four questions for which we detected a significant effect, we can see there is more density at p-values less than 0.05 compared to the null of no effect (uniform distribution of p-values). Values less than 0.05 are highlighted in red.\\

\begin{figure}[htp]
    \centering
    \includegraphics[width=5in,height=5in]{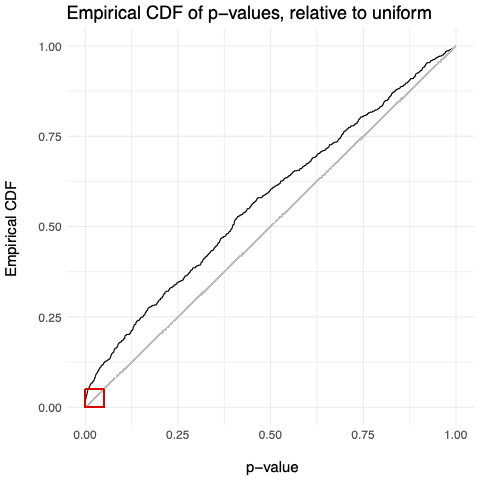}
    \caption*{Figure S1. Empirical CDF of treatment effect p-values across experiments.}
    \label{fig:heterogeneity1}
\end{figure}

\begin{figure}
\begin{tabular}{cccc}
\subfloat{\includegraphics[width = 3in]{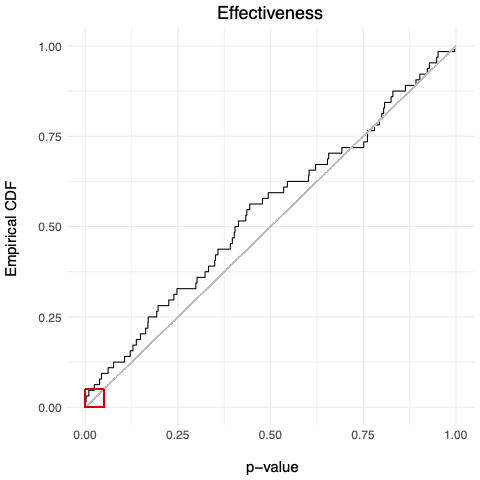}} &
\subfloat{\includegraphics[width = 3in]{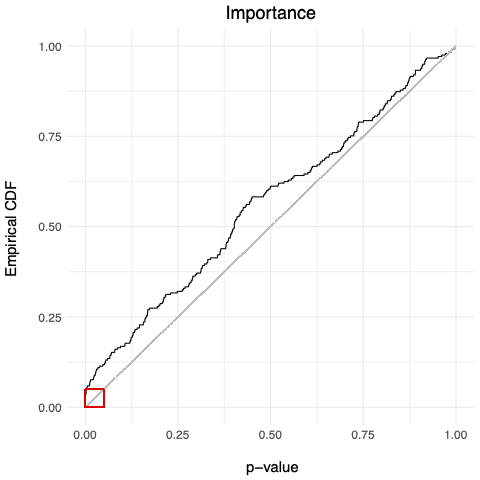}} \\
\subfloat{\includegraphics[width = 3in]{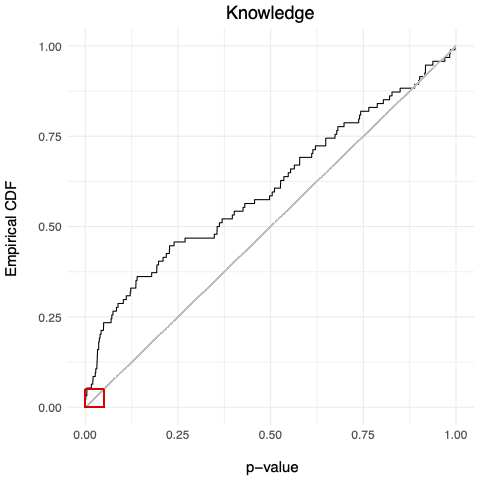}} &
\subfloat{\includegraphics[width = 3in]{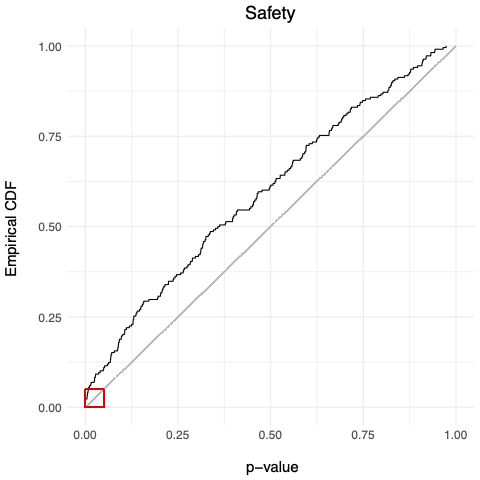}}
\end{tabular}
\end{figure}

\begin{figure}
\begin{tabular}{cccc}
\subfloat{\includegraphics[width = 3in]{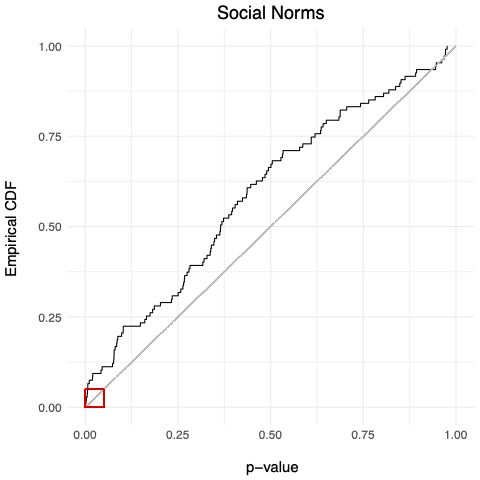}} &
\subfloat{\includegraphics[width = 3in]{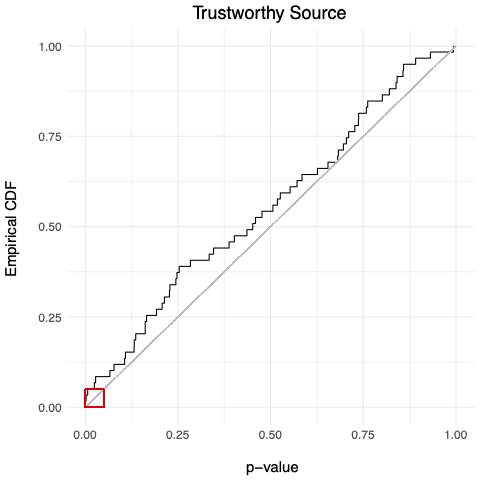}} \\
\subfloat{\includegraphics[width = 3in]{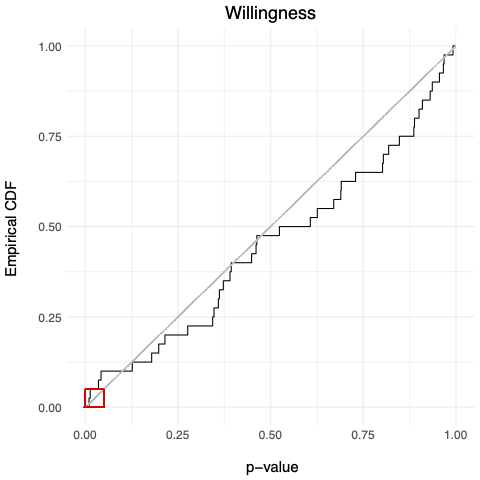}} 
\end{tabular}
\caption*{Figure S2. Empirical CDFs of treatment effect p-values across experiments, by outcome. Note how the four questions we observe significant effects for (Importance, Knowledge, Safety, Social Norms) each have CDFs with substantial mass above the uniform distribution for $p<0.05$.}
\end{figure}

\pagebreak
\subsection*{Alternative Estimates of Cost Effectiveness}

There are multiple ways one could think about trying to quantify an aggregate effect across these campaigns in terms of the total number of influenced people and the cost per influenced person. Our preferred approach, as detailed in the main text, is to treat the set of campaigns effectively as one aggregated campaign. Specifically, we can look at the total number of unique users who were exposed to these campaigns and scale that by our treatment effect estimate to get a back of the envelope sense of how many people may have been influenced. To derive an estimate of cost per influenced person, we can then divide total spend for those campaigns by this number. 

As a numerical example, to derive the overall estimate of cost per influenced person, we take \$39.4 million (total spend) and divide it by .0055 (ATE estimate) times 2.1 billion (unique reach), which gives \$3.41. We note at the individual campaign level, this procedure is a common industry practice and how advertisers are used to estimating total campaign effects.

One issue with this approach is that it ignores the fact that there were users who were in multiple campaigns. If a campaign changed a person's mind about Willingness and a separate campaign changed her mind about Importance, some would argue that should count as two influenced events instead of one. This would be an argument for scaling up each individual campaign by its reach, and then summing those estimates across all campaigns; here, instead of influenced `people' arguably influenced `opinions' would be more appropriate. In Table S4 we report out results from this method as well; unsurprisingly, the cost per influenced outcome droops substantially.

Finally, we note because there are individuals who were exposed to multiple campaigns, when we sum the total number of influenced people across columns it will not add up to the overall estimate (the overall estimate will be smaller). This is because a person will appear in as many columns as they were exposed to campaigns that asked a given question; in the overall calculation we are only counting such people once, weighed by the average treatment effect estimate.

\begin{table}[htbp]
\caption*{Table S4. Alternative Approaches to Calculating Cost per Outcome Influenced.}
\resizebox{\columnwidth}{!}{\begin{tabular}{l c c c c c c c c}
\hline
Category & Effectiveness & Importance & Knowledge & Safety & \makecell{Social \\Norms} & \makecell{Trustworthy \\ Source} & Willingness & Overall\\[0.1ex] 
\hline\hline
\\
\multicolumn{9}{l}{\textbf{Main results}}\\
\\
Treatment Coefficient 	& 0.0045 		& 0.0043*** 	& 0.0123*** 	& 0.0062*** 	& 0.0081*** 	& 0.0012 		& 0.0010 		& 0.0055***\\
                      			& (0.0029) 	& (0.0012) 	& (0.0025) 	& (0.0016) 	& (0.0024) 	& (0.0027) 	& (0.0042) 	& (0.0008)\\
\\
p-value 				& 0.114		& 0.0004 		& 5e-7 		& 8e-5 		& 0.0006 		& 0.639 		& 0.807 		& 2e-13\\
\\
\hline
\\
\# Total Reach & 791,144,678 & 2,985,622,768 & 123,2041,831 & 2,595,117,781 & 1,312,200,394 & 561,139,998 & 1,091,904,828 & 4,842,392,644\\
\\
\# Unique People Reached & 623,509,126 & 1,424,954,066 & 859,760,237 & 1,472,804,702 & 828,296,583 & 411,859,697 & 835,456,995 & 2,094,285,077\\
\\
\# Total Spend & \$7,072,815 & \$14,676,490 & \$8,294,165 & \$28,750,458 & \$6,673,498 & \$6,146,421 & \$15,707,139 & \$39,443,567\\
\\
{\bf Preferred Approach} &&&&&&&& \\
\\
\hspace{.6cm} \emph{Number of Influenced People} & 2,815,542 & 6,128,713 & 10,596,845 & 9,061,267 & 6,723,167 & 511,803 & 854,596 & 11,583,543\\
\\
\hspace{.6cm} \emph{Cost per Influenced Outcome} & \$2.51 & \$2.39 & \$0.78 & \$3.17 & \$0.99 & \$12.01 & \$18.38 & \$3.41\\
\\
{\bf Alternative Approach} &&&&&&&& \\
\\
\hspace{.6cm} \emph{Number of Influenced People} & 3,572,524 & 12,841,133 & 15,185,346 & 15,966,173 & 10,650,947 & 697,309 & 1,116,919 & 26,783,394\\
\\
\hspace{.6cm} \emph{Cost per Influenced Outcome} & \$1.98 & \$1.14 & \$0.55 & \$1.80 & \$0.63 & \$8.81 & \$14.06 & \$1.47\\
\\
\hline
\end{tabular}}
\end{table}

\pagebreak
\subsection*{Characteristics of Survey Respondents Across Groups}

Below we compare how the sample of survey respondents may differ (i) across treatment and control, and (ii) versus the overall population who was exposed to the campaign. The aim of the first analysis (Table S5) is to mitigate concerns that differences in survey outcomes between the groups may be due to a factor other than the treatment ad (e.g., differential response rates). If we observed imbalances in any observable that may be cause for concern given our exposure randomization. Due to data retention limitations, we unfortunately could not recover this data from the same set of experiments in our sample; however, the data below are pulled from 15 COVID-related Brand Lift Studies that were run more recently (these are from the same collection we analyze, just more recent). We can see that at least in this sample, the demographics are not significantly different other than for age, which we control for.

\begin{table}[htbp]
\caption*{Table S5. Balance of demographics across treatment and control for survey respondents. The p-value is from a two-sided t-test comparing the difference in means across treatment and control; the adjusted p-value uses the same Benjamini-Hochberg procedure as in the main text.}
\begin{tabular}{l r r r r}
\hline
Demographic & \makecell{$\bar{x}_{control}-\bar{x}_{test}$} &  \makecell{Std. Error\\ of Diff.} & p-value & Adj. p-value\\[0.1ex] 
\hline\hline
\\
Age	(years) & -0.71	& 0.23& 	0.002 & 0.0374\\
Gender& 	-0.01& 	0.01& 	0.389 & 1.000\\
Friend Count & 	0.01	&15.81	&0.9995&	1.000\\
\# Friendships Initiated &	-2.96 &	7.58	& 0.6965 &	1.000 \\
Subscription Count &	-6.75	&13.82&	0.6251	& 1.000\\
Subscriber Count& 	-7.86&	11.06&	0.4774&	1.000\\
Long term user & 	0.00& 	0.01& 	0.6196 & 1.000\\
Android user& 	-0.01&	0.01&	0.1864&	1.000\\
Profile Photo Present& 	0.00& 	0.00& 	0.9974 & 1.000\\
Birthday Privacy setting & 	0.00& 	0.00& 	0.5651 & 1.000\\
Contact Email Confirmed	& 0.00& 	0.00& 	0.2604 & 1.000\\
\# Days Active on Platform last 7d	& 0.00& 	0.00& 	0.9724 & 1.000\\
Month of birth& 	-0.01& 	0.05& 	0.8029 & 1.000\\
Single & 	-0.01& 	0.00& 	0.1341 & 1.000\\
US Based& 	0.00& 	0.01& 	0.4413 & 1.000\\
\hline \\
\end{tabular}

\footnotesize{Note: Long term users are defined as having been on the site more than five years. Birthday privacy setting is a dummy indicating whether other users on platform can see it or not. These are common, well-populated descriptives of users, either from self-report on their profiles or log data.}\\

\end{table}

The aim of the second analysis (Table S6) is to address representativeness of our sample, and how well we think it might generalize to the broader target population. For this, we take a sample of survey respondents that has been post-stratified by age bucket and gender category back to the population that the campaign reached (as we do in the main text) and compare the reweighted demographics versus those of the target population. To the extent that the reweighted demographics match those of the target population, we can feel more comfortable in how generalizable our survey-based answers are to the broader population. We note here that in this comparison it is less paramount to see close matching along all demographics; indeed, conditional only on age and gender, we would still expect to see differences across populations. To the extent that the two may differ along a demographic that we think may be correlated with our treatment variable represents a concern, however. 

Again due to data retention issues, we cannot use data from our broader study, but report results here for a recent COVID Brand Lift Study where we were able to obtain the relevant data. We can see below that along a relatively large number of demographics the reweighted sample and population means are not significantly different. 

The variable which has the largest difference between the groups and which we may be concerned about as correlating with our treatment effect is the Android user indicator. We note that rerunning our main specification for Android users only, we do not detect a significant difference versus our current estimated treatment effect ($p$=0.79).\\

\begin{table}[htbp]
\caption*{Table S6. Balance of demographics across campaign population and age bracket and gender post-stratification weighted survey responses. The last column reports the p-value for a two-sided t-test comparing the difference in means across treatment and control.}
\begin{tabular}{l r r r r}
\hline
Demographic & \makecell{$\bar{x}_{pop}-\bar{x}_{sample}$} &  \makecell{Std. Error\\ of Diff.} & p-value & Adj. p-value\\[0.1ex] 
\hline\hline
\\
Age	(years) & 0.03	& 0.10 & 0.779 & 1.000 \\
Gender & 	-0.00 &	0.00	&0.617 &	1.000\\
Friend Count& 	47.34	& 30.43 &	0.120 &	1.000\\
\# Friendships Initiated	& 27.16	& 14.38 &	0.059	&0.589\\
Subscription Count& 	-83.10 &	27.56	& 0.002	& 0.036 \\
Subscriber Count & 	-28.20	& 26.30 &	0.284 &	1.000\\
Long term user& 	0.05	 & 0.02 &	0.003 &	0.043\\
Android user& 	0.23	& 0.01 &	0.000 &	0.000\\
Profile Photo Present& 	-0.01 & 	0.00 &	0.031 &	0.366\\
Birthday Privacy setting (=not visible) & 	-0.01	 & 0.01	& 0.317 &	1.000\\
Contact Email Confirmed	& -0.01	& 0.01 &	0.302	&1.000\\
\# Days Active on Platform last 7d	& 	-0.02 & 	0.01	& 0.038 &	0.416\\
Month of birth& 	-0.14	 & 0.13 &	0.267 &	1.000\\
Single & 	-0.01 & 	0.01 & 	0.419 &  1.000\\
US Based& 	-0.00& 	0.00& 	0.288 & 1.000\\
\hline \\
\end{tabular}
\end{table}

\noindent \\

\pagebreak
\subsection*{Survey Details}

We note that exact wording within a language was largely the same across experiments, but advertisers did have autonomy to tweak if they wanted. Questions for Knowledge, Social Norms, and Trustworthy Source had a bit more variation than the others, but the questions still targeted the same concept. Questions were translated into the languages specified by a user's settings. Core questions are noted below, with responses and coding as desired/not in parentheses after each response: \\

\singlespacing

\noindent {\bf Importance}
\begin{itemize}
    \item How important do you feel a vaccine is to prevent the spread of COVID-19?
    \begin{itemize}
    \item Very important (1)
    \item Somewhat important (1)
    \item Barely important (0)
    \item Not important (0)
    \item I don't know (0)
\end{itemize}
\end{itemize}

\noindent \\
\noindent {\bf Safety}
\begin{itemize}
    \item How safe do you think a COVID-19 vaccine is for people like you?
    \begin{itemize}
    \item Very safe (1)
    \item Somewhat safe (1)
    \item Barely safe (0)
    \item Not safe (0)
    \item I don't know (0)
\end{itemize}
\end{itemize}

\noindent \\
\noindent {\bf Willingness}
\begin{itemize}
    \item How likely are you to get vaccinated for COVID-19 when the vaccine is available to you?
    \begin{itemize}
    \item Very likely (1)
    \item Somewhat likely (1)
    \item Somewhat unlikely (0)
    \item Very unlikely safe (0)
    \item I don't know/I already got vaccinated (0)
\end{itemize}
\end{itemize}

\noindent \\
\noindent {\bf Effectiveness}
\begin{itemize}
    \item How effective do you think the COVID-19 vaccination is in preventing COVID-19?
    \begin{itemize}
    \item Very effective (1)
    \item Somewhat effective (1)
    \item Barely effective (0)
    \item Not effective (0)
    \item I don't know (0)
\end{itemize}
\end{itemize}

\noindent \\
\noindent {\bf Knowledge}
\begin{itemize}
    \item Do you know where people in your local community can go to get a COVID-19 vaccine?
    \begin{itemize}
    \item Yes (1)
    \item No (0)
    \item I don't know (0)
\end{itemize}
\end{itemize}

\noindent \\
\noindent {\bf Social Norms}
\begin{itemize}
    \item When you think of most people whose opinion you value, how much would they approve of people getting a COVID-19 vaccine?
    \begin{itemize}
    \item Definitely approve (1)
    \item Mostly approve (1)
    \item Somewhat approve (0)
    \item Not at all approve (0)
    \item I don't know (0)
\end{itemize}
\end{itemize}

\noindent \\
\noindent {\bf Trustworthy Source}
\begin{itemize}
    \item Do you agree or disagree that [advertiser name] is a trustworthy source of COVID-19 vaccine facts and information?
    \begin{itemize}
    \item Strongly agree (1)
    \item Somewhat agree (1)
    \item Somewhat disagree (0)
    \item Strongly disagree (0)
    \item I don't know (0)
    \end{itemize}
\end{itemize}

\end{document}